# Photoneutron reactions on $^{93}$Nb at $E_{\gamma max}$ = 33 ÷ 93 MeV

A. N. Vodin[1], O. S. Deiev[1], V. Yu. Korda[2], I. S. Timchenko[1], S. N. Olejnik[1],
N. I. Aizatsky[1], A. S. Kachan[1], L. P. Korda[1], E. L. Kuplennikov[1],
V. A. Kushnir[1], V. V. Mitrochenko[1], and S. A. Perezhogin[1]

[1]*NSC Kharkov Institute of Physics and Technology, National Academy of Sciences of Ukraine, 1 Akademicheskaya Street, UA-61108 Kharkov, Ukraine and*
[2]*Institute of Electrophysics and Radiation Technologies, National Academy of Sciences of Ukraine, 28 Chernyshevsky Street, Post Office Box 8812, UA-61002 Kharkov, Ukraine*

The bremsstrahlung flux-averaged cross-sections for the photoneutron reactions $^{93}$Nb($\gamma$,$x$n; $x$ = 1 - 5)$^{(93-x)m,g}$Nb were measured in the range of boundary energies of bremsstrahlung $\gamma$-quanta $E_{\gamma max}$ = 33 ÷ 93 MeV with a step $\Delta E_{\gamma max}$ ≈ 2 MeV. The isomeric ratios of the average cross-sections of the products of the reactions $^{93}$Nb($\gamma$,4n)$^{89m,g}$Nb and $^{93}$Nb($\gamma$,5n)$^{88m,g}$Nb were determined in the energy ranges $E_{\gamma max}$ = 50 ÷ 93 and 70 ÷ 93 MeV, respectively. The experiments were carried out on the beam of the linear electron accelerator LUE-40 of the Science and Research Establishment (SRE) "Accelerator" at National Science Center Kharkov Institute of Physics and Technology (NSC KIPT) using the method of induced activity. Calculations of the cross-sections, average cross-sections, and isomeric ratios of the reaction products were performed using the TALYS1.9 code with default parameters and the GEANT4 code. The tendency of a more successful description of the average cross-sections of photoneutron reactions with the formation of final odd-even Nb nuclei than odd-odd Nb nuclei is revealed. The experimental average cross-sections for the reactions ($\gamma$,2n) and ($\gamma$,4n) are in good agreement with theory, while in the case of reactions ($\gamma$,n), ($\gamma$,3n), and ($\gamma$,5n), some discrepancies are observed. The results obtained for the reactions ($\gamma$,n), ($\gamma$,3n) and ($\gamma$,4n) are in satisfactory agreement with the known literature data. The average cross-sections for the reactions ($\gamma$,2n) and ($\gamma$,5n) and the isomeric ratios of the reaction products $^{93}$Nb($\gamma$,5n)$^{88m,g}$Nb were measured for the first time.

## I. INTRODUCTION

The overwhelming majority of data on the energy dependence of the cross-sections for photonuclear reactions were obtained in the study of giant dipole resonance (GDR) in experiments performed using bremsstrahlung $\gamma$-radiation and quasi-monoenergetic photons [1-5] at $\gamma$-ray energies up to ~ 30 MeV. At the same time, the study of photodisintegration of nuclei in the energy range above the GDR and up to the pion production threshold ($E_{th}$ ≈ 145 MeV) is of undoubted interest due to the change in the mechanism of interaction of photons with nuclei in this energy region. An important direction of research is opening up that will make it possible to obtain fundamental information on the competition between two mechanisms of photodisintegration of nuclei in this energy range - through the excitation of GDR and quasi-deuteron photoabsorption. However, the general lack of detailed experimental data severely limits both the general understanding of the processes of interaction of $\gamma$-quanta with nuclei in the energy range ~ 30 ÷ 145 MeV, and the possibilities for testing model approaches [6-8]. Undoubtedly, the main reason hindering the further development of such studies is the absence of intense sources of quasi-monochromatic $\gamma$-quanta, which is fundamentally important for relatively small cross-sections for multiparticle photonuclear reactions in the indicated energy range. In this context, linear electron accelerators with smooth energy tuning can be considered as alternative sources of γ-radiation capable of generating intense beams of medium and high-energy bremsstrahlung γ-quanta. The use of bremsstrahlung beams with a continuous spectrum, taking into account the development of computer data processing methods, is a good alternative to quasi-monochromatic γ-quanta for carrying out experiments on photodisintegration of nuclei in the considered energy range of γ-quanta.

At present, data on the cross-sections for photonuclear reactions with the yield of multiple nucleons are in demand in a variety of applications, which also stimulates interest in such studies. For example, when creating nuclear facilities based on subcritical systems, the so-called accelerator driven systems (ADS), controlled by electron or proton accelerators. Currently, ADS systems are considered promising installations for the disposal of radioactive waste from nuclear power [9], as well as potential sources for the production of electricity [10, 11]. Since such installations operate in a subcritical mode, it is very important to know the flux density of additional neutrons that are formed as a result of multiparticle photoneutron reactions on the nuclei of structural materials (Al, Fe, Zr, Nb, etc.) in the ADS core. This will make it possible to more correctly estimate the total neutron flux in the core of the subcritical assembly, which, in turn, is associated with the safe operation of such nuclear facilities.

The cross-sections for photoneutron reactions are in demand in studies of the properties of quark-gluon plasma, which are currently being actively carried out in colliding beams of relativistic heavy nuclei at the colliders RHIC (Brookhaven National Laboratories, USA) and LHC (CERN) [12-14]. In the very first experiments it was found that electromagnetic dissociation (EMD) of ultrarelativistic nuclei significantly affects the lifetime of beams [15], and secondary ions exert a local thermal effect on superconducting magnets [16]. It is expected that in collisions of light and medium nuclei ($^{40}$Ar, $^{78,84}$Kr, $^{129}$Xe) these effects can be weakened [17-19]. However, the planning of such experiments is associated with a reliable estimate of the EMD cross-sections, the calculation of which depends on the use of various approximations of the total cross-sections for photon absorption by nuclei. In this regard, for the further development of the theory of photonuclear reactions, experimental data on the cross-sections for the absorption of photons by nuclei obtained in a wide range of energies of γ-quanta, including in the energy range under consideration, are required.

The interest in studying photonuclear reactions on the $^{93}$Nb nucleus is due to several reasons. Photodisintegration reactions of $^{93}$Nb are accompanied by the formation of final nuclei in the mass region $A \sim 90$, where shell effects associated with filling the neutron shell with $N = 50$ and the proton subshell with $Z = 40$, which are of interest for the static theory of nuclear reactions, can be clearly manifested. In addition, niobium is a part of zirconium alloys, which are used in the manufacture of primary cladding of fuel rods in the countries of the former USSR and Canada. For example, fuel rod claddings are made of such an alloy, which are used at the NSC KIPT neutron source (built jointly with Argonne National Laboratory, USA) based on a subcritical assembly with low-enriched uranium fuel (~ 20%) controlled by an electron accelerator with an energy of 100 MeV and the beam power of 100 kW [10, 11].

Sufficiently complete data on the cross-sections for the reactions (γ,n) and (γ,2n) on $^{93}$Nb were obtained in the GDR region on a beam of quasi-monochromatic annihilation photons in the energy range 8 ÷ 25 MeV using the direct neutron detection method [20]. In the same energy range, data are known on the bremsstrahlung flux-averaged cross-sections for the reaction $^{93}$Nb(γ,n)$^{92m}$Nb at the boundary energies of bremsstrahlung γ-quanta $E_{\gamma max}$ = 10 and 12.5 MeV [21] and at $E_{\gamma max}$ = 12, 14, and 16 MeV [22] obtained using the induced activity method.

The first experiments on photodisintegration of the $^{93}$Nb nucleus in the energy region above the GDR maximum were carried out in Ref. [23], in which the average cross-sections for

electro- and photonuclear reactions of the type ($\gamma$,$x$n$y$p: $x$ = 1 - 6, $y$ = 1, 2) were measured in the energy range $E_{\gamma max}$ = 100 ÷ 275 MeV (step $\Delta E_{\gamma max}$ = 25 MeV) using the method of induced activity. For the reaction ($\gamma$,n), the authors of Ref. [23] obtained good agreement between the experimental cross-sections and theoretical calculations predicted by the cascade-evaporation model and the Rudstam formula [24]. At the same time, a significant excess of experiment over theory for the reaction ($\gamma$,3n) was revealed, which is explained by the influence of filling in this region of masses of neutron shell nuclei with $N$ = 50. In Ref. [25], using the activation method, the integral cross-sections $\sigma_{int}$ = 1.2 MeV·b of the reaction $^{93}$Nb($\gamma$,$\alpha$)$^{89m}$Y was measured at $E_{\gamma max}$ = 25 MeV, and it was found that the contribution of direct processes to the cross-sections of this reaction is small compared to reaction ($\gamma$,p). In a study by Aliev et al. [26] the ratio of the yields of the reactions Y($\gamma$,n)/Y($\gamma$,3n) = 63±1 at $E_{\gamma max}$ = 50 MeV was obtained and the corresponding theoretical estimate was made - 97, based on the model of evaporation. In Ref. [27], the contribution of the average cross-sections for the formation of $^{92}$Nb in the metastable-state to the average total cross-section for the reaction $^{93}$Nb($\gamma$,n)$^{92}$Nb at $E_{\gamma max}$ = 32 MeV was experimentally determined, which turned out to be 55.2%. More detailed data on the average cross-sections for reactions ($\gamma$,$x$n; $x$ = 1, 3, 4) on $^{93}$Nb were obtained by Naik et al. [22] at the boundary bremsstrahlung energies of 45 - 70 MeV (step $\Delta E_{\gamma max}$ = 5 and 10 MeV) using the method of induced activity. The authors of Ref. [22] noted good agreement between the experimental average cross-sections for the reactions ($\gamma$,n), ($\gamma$,3n), and ($\gamma$,4n) with the calculated values obtained using the TALYS 1.4 code. In Ref. [28], the first attempt was made to obtain data on the isomeric ratios of the reaction products $^{93}$Nb($\gamma$,4n)$^{89m,g}$Nb at $E_{\gamma max}$ = 60 and 70 MeV. Naik et al. [29] continued these experiments at the boundary energies of the bremsstrahlung $\gamma$-radiation $E_{\gamma max}$ = 45 - 70 MeV, however, the data on the isomeric ratios of the yields turned out to be much higher than the calculated values obtained using the TALYS 1.4 code. In the studies by Demira and Çetin [30], the radiative strength function and the cross-section for the $^{93}$Nb($\gamma$,n)$^{92}$Nb reaction were calculated using the TALYS 1.6 code [31], as well as the effect of pre-equilibrium processes on the reaction cross-sections in the region $E_\gamma$ = 8 ÷ 200 MeV was studied. Good agreement was noted between the calculated values obtained and the experimental data known at that time. This result, according to the authors, suggests that the TALYS code with default parameters can be used to calculate photonuclear cross-sections with good accuracy.

Summarizing the analysis of the known experimental data on the photodisintegration of $^{93}$Nb in the region of $\gamma$-quanta energies above 30 MeV, it should be noted that they are extremely discrete and local. It will be fair to say that a similar situation has developed for other nuclei, despite the many years of efforts made by experimenters to solve this problem [32, 33].

The aim of this work is to obtain new and more detailed experimental data on the average cross-sections and isomeric ratios of the products of photoneutron reactions ($\gamma$,$x$n; $x$ = 1 - 5) on the $^{93}$Nb nucleus. The studies were carried out in a wide range of boundary energies of bremsstrahlung photons $E_{\gamma max}$ = 33 ÷ 93 MeV with an arbitrary step $\Delta E_{\gamma max} \approx$ 2 MeV using the method of induced activity and the $^{100}$Mo($\gamma$,n)$^{99}$Mo monitor reaction. All targets for all $E_{\gamma max}$ were irradiated during the time $t_{irr}$ = 30 min. This made it possible to obtain information on the average cross-sections for the production of final nuclei in photoneutron reactions on $^{93}$Nb with lifetimes from several minutes to several months with good statistics. The experimental results are compared with the calculations performed within the framework of the statistical model implemented in the open program code TALYS1.9 [34, 35]. The calculations in TALYS1.9 were carried out with default parameters in order to describe from a unified standpoint the energy dependence of isomeric ratios and cross-sections for the observed reactions on the $^{93}$Nb nucleus in the studied range of $\gamma$-ray energies.

## II. EXPERIMENTAL PROCEDURE

### A. Accelerator complex

The experiments were carried out on the beam of bremsstrahlung γ-radiation of the linear accelerator of electrons LUE-40 SRE "Accelerator" NSC KIPT [36] using the well-known method of induced activity of the final nucleus of the reaction. In LUE-40, the main acceleration of electrons is carried out in two accelerating sections with a quasi-constant structure, which allows to quickly adjust the particle energy by changing the power and phase of the microwave power supply of the second section. This technique of tuning the accelerator makes it possible to ensure a smooth change in the energy of accelerated electrons in the range $E_e = 30 \div 100$ MeV at an average beam current $I_e \sim 3$ μA. In this case, the full width at half maximum (FWHM) of the energy spectrum of electrons is $\Delta E_e/E_e \approx 1$ % at a pulse repetition rate of 50 Hz and a pulse length of 10 μs. The block diagram of this experiment is shown in Fig. 1.

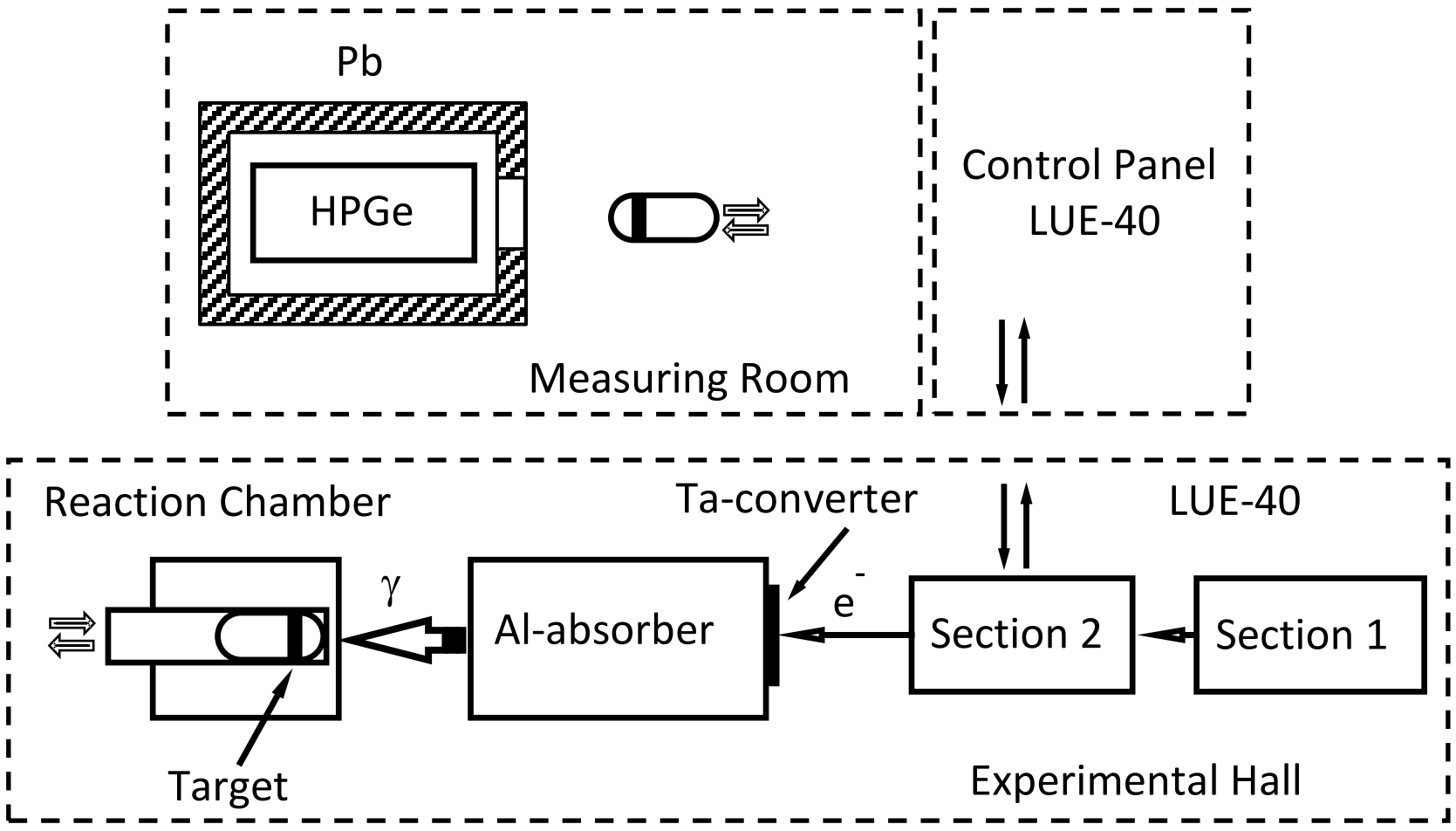

FIG. 1. The block diagram of the experiment.

Bremsstrahlung was generated when a pulsed electron beam passed through a tantalum (Ta) converter 10×15 mm in sizes and 1.05 mm thick. The Ta-converter was attached to a pure aluminum (Al) cylinder with dimensions Ø100×100 or 150 mm, which was used as an absorber of electrons that passed through the converter. The electron beam incident on the converter had a transverse size of ≤ 4 mm.

### B. Target

For the experiments, 29 targets were made from $^{93}$Nb and the same number from natural $^{nat}$Mo with chemical purities of 99.99% and 99.9%, respectively. The targets were thin disks 8 mm in diameter and $m \sim 80$ mg for $^{93}$Nb and $m \sim 60$ mg for $^{nat}$Mo. For all samples, the accuracy of determining their mass was $\Delta m/m \leq 0.01\%$. In the experiment, stacks of $^{93}$Nb + $^{nat}$Mo targets were irradiated for $t_{irr} = 30$ min at all electron energies. After changing the boundary bremsstrahlung energy, a new pair of $^{93}$Nb and $^{nat}$Mo targets was installed in the reaction chamber with an accuracy of displacement of its center relative to the geometric axis of the electron beam no worse than ± 0.5 mm. Targets were delivered to the reaction chamber using a pneumatic transport system in a special aluminum capsule. The position of the capsule in the reaction chamber was fixed with a photodiode sensor. After the irradiation session, both targets were moved to the measurement room, in which the activated $^{93}$Nb and $^{nat}$Mo samples were removed from the capsule and alternately placed on an adjustment table in front of the HPGe-detector. The pneumatic transport system ensured continuous operation of the accelerator in a given mode and determined the minimum target cooling time $t_{cool}$ = 30 s, which, in principle, made it possible in the present experiment to measure γ-radiation of activated samples with a half-life of up to one minute.

### C. γ-ray spectrum and data analysis

Two sessions of measurements of the γ-radiation spectra of final nuclei formed in photoneutron reactions on $^{93}$Nb and $^{nat}$Mo were carried out. In the first measurement session, the activated $^{93}$Nb samples were cooled for $t_{cool}$ = 1.5 min and then their γ-emission spectra were measured for $t_{meas}$ = 30 min. After that, the spectra of γ-radiation of nuclides formed during the photodisintegration of $^{nat}$Mo were measured in a time $t_{meas}$ = 30 min. In the second session with $t_{cool}$ = 60 d and $t_{meas}$ = 24 h, the spectra of γ-radiation only for the nuclides formed in the (γ,n) and (γ,2n) reactions on $^{93}$Nb were measured. Thus, information was obtained on the yields of reactions with the formation of Nb isotopes with half-lives from several minutes to several months.

The measurements were performed using a Canberra HPGe-detector with a FWHM resolution $\Delta E_\gamma$ = 1.8 keV for the γ-line with $E_\gamma$ = 1332 keV $^{60}$Co and an efficiency of 20%, relative to the NaI(Tl)-detector with dimensions Ø300×300 mm. The HPGe-detector was placed in a special lead tube, which significantly improved the background measurement conditions. To calibrate the spectrometric channel in terms of energy and absolute efficiency of registration of γ-quanta ε, reference sources of γ-radiation $^{22}$Na, $^{60}$Co, $^{133}$Ba, $^{137}$Cs, $^{152}$Eu, and $^{241}$Am were used. When collecting γ-spectra, the dead time of the spectrometric path did not exceed 3 – 5 %. The loading of the spectrometer was controlled by adjusting the distance $D$ between the activated sample and the HPGe-detector. In this regard, the value of ε was determined for various energies of γ-quanta for several values of $D$ and was further approximated by the analytical curve in the form $\ln\varepsilon = \sum_i^n a_i \times (\ln E_\gamma)^i$ proposed in Ref. [37].

Figure 2 shows a typical spectrum of γ-radiation of an activated $^{93}$Nb target after its irradiation with a flux of bremsstrahlung γ-quanta at $E_{\gamma max}$ = 85.5 MeV for $t_{irr}$ = 30 min and $t_{meas}$ = 30 min. For convenience, the spectrum is presented in the form of two parts: for $0 \le E_\gamma \le 1600$ keV [Fig. 2(a)] and for $1600 \le E_\gamma \le 3200$ keV [Fig. 2(b)]. Inset in Fig. 2(a) shows a fragment of the spectrum in the vicinity of the γ-line with $E_\gamma$ = 1204.8 keV obtained for $t_{irr}$ = 30 min, $t_{cool}$ = 60 d, and $t_{meas}$ = 24 h. The spectrum contains γ-lines corresponding to the emission of final product nuclei formed as a result of photodisintegration of $^{93}$Nb. In some cases, the γ-spectra were recorded several times at certain time intervals in order to clarify the half-life of the daughter nucleus formed in the corresponding channel of the photonuclear reaction. After determining the sum of γ-quanta in the peak of total absorption, the experimental activity of the target along the selected decay channel at the time of the end of irradiation was calculated, which was subsequently used to determine the average cross-sections of the corresponding reaction. The Canberra Genie-2000 software [38] was used to analyze the γ-ray spectra.

The daughter nucleus was identified by the energies of γ-lines, taking into account their tabular values for the absolute intensities and half-lives, the data on which were taken from Refs. [39-41]. Table I shows the main spectroscopic characteristics of radionuclides formed as a result of photoneutron reactions on $^{93}$Nb and $^{100}$Mo nuclei. The energies and intensities of the γ-lines (only the most intense γ-lines are indicated), which were directly used in the analysis of the data of the present experiment, are highlighted in Table I in bold.

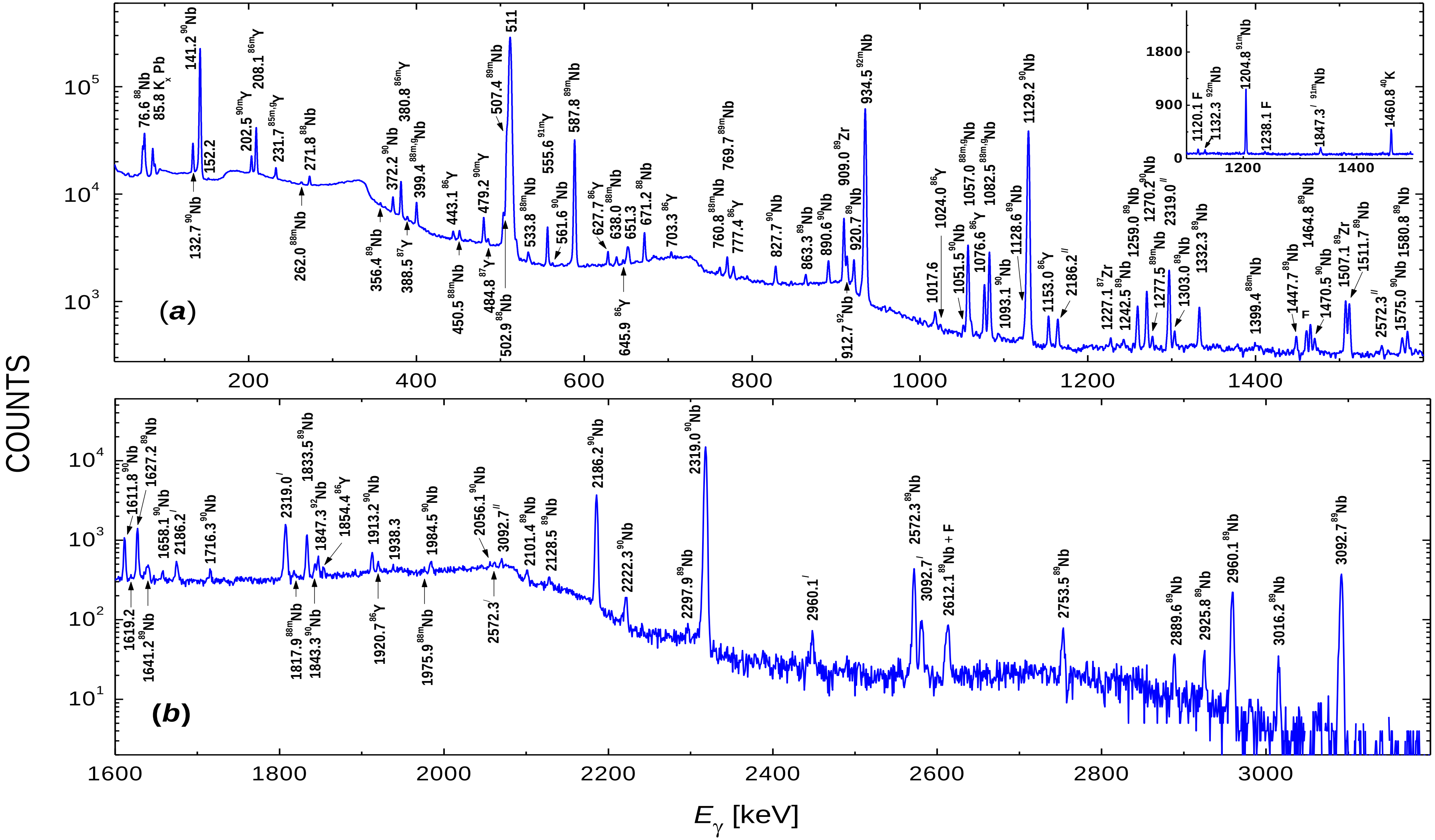

FIG.2. Spectrum of γ-radiation of a $^{93}$Nb target weighing 80 mg after its irradiation with a flux of bremsstrahlung photons at boundary energy $E_{\gamma max}$ = 85.5 MeV for 30 min and a measurement time of 30 min. (a) 0 ≤ $E_\gamma$ ≤ 1600 keV. (b) 1600 ≤ $E_\gamma$ ≤ 3200 keV. Inset - a fragment of the spectrum in the vicinity of the γ-line with $E_\gamma$ = 1204.8 keV obtained for irradiation time 30 min, cooling time 60 d, and measurement time 24 h. Letter 'F' marks natural γ-background, superscript '/' marks the first escape peak, and superscript '//' marks the second escape peak.

TABLE I. Nuclear spectroscopic data of the radionuclides from the $^{93}$Nb(γ,$x$n; $x$ = 1-5)$^{(93-x)m,g}$Nb and $^{100}$Mo(γ,n)$^{99}$Mo reactions. $E_{th}$ is the threshold of the corresponding reaction. $J$, π, and $T_{1/2}$ are spin, parity, and the half-life of the state. The energies $E_γ$ and intensities $I_γ$ of the γ-lines (only the most intense γ-lines are indicated), which were directly used in the analysis of the data of the present experiment, are shown in bold.

| Nuclear reaction | $E_{th}$, MeV | Reaction product | $J^π$ | $T_{1/2}$ | $E_γ$, keV | $I_γ$, % |
|---|---|---|---|---|---|---|
| $^{93}$Nb(γ,n)$^{92}$Nb | 8.45 | $^{92m}$Nb | $(2)^+$ | 10.15(2) d | **934.46** | **99** |
| | 8.31 | $^{92g}$Nb | $(7)^+$ | 3.47(24)·10$^7$ y | 561.03 | 100 |
| | | | | | 934.46 | 100 |
| $^{93}$Nb(γ,2n)$^{91}$Nb | 16.83 | $^{91m}$Nb | $1/2^-$ | 60.86(22) d | 104.62 | 0.539(19) |
| | | | | | **1204.77** | **2.9** |
| | 16.72 | $^{91g}$Nb | $9/2^+$ | 680(13) y | - | - |
| $^{93}$Nb(γ,3n)$^{90}$Nb | 28.90 | $^{90m}$Nb | $4^-$ | 18.81(6) s | 122.370 | 64 |
| | 28.77 | $^{90g}$Nb | $8^+$ | 14.60(5) h | 141.18 | 66.8(7) |
| | | | | | **1129.22** | **92.7(4)** |
| | | | | | **2186.24** | **17.96(16)** |
| | | | | | **2318.97** | **82.03(16)** |
| $^{93}$Nb(γ,4n)$^{89}$Nb | 38.85 | $^{89m}$Nb | $(1/2^-)$ | 1.18(10) h | 587.83 | 100 |
| | | | | | **769.69** | **6.5(6)** |
| | 38.85 | $^{89g}$Nb | $(9/2^+)$ | 1.9(2) h | **1511.7** | **1.87(14)** |
| | | | | | **2572.3** | **2.58(20)** |
| | | | | | **2960.1** | **1.70(17)** |
| | | | | | **3092.7** | **3.0(3)** |
| $^{93}$Nb(γ,5n)$^{88}$Nb | 51.41 | $^{88m}$Nb | $(4^-)$ | 7.8(1) min | **450.52** | **26.6(8)** |
| | | | | | **638.00** | **26.2(8)** |
| | 51.41 | $^{88g}$Nb | $(8^+)$ | 14.5(1) min | **271.80** | **30.1(9)** |
| | | | | | **671.20** | **64(2)** |
| $^{100}$Mo(γ,n)$^{99}$Mo | 8.29 | $^{99g}$Mo | $1/2^+$ | 65.94(1) h | **140.51** | **89.43(23)** |
| | | | | | **739.50** | **12.13(12)** |

**D. Calculation of the average cross-sections for photoneutron reactions**

The calculation of the cross-sections σ($E$) for photoneutron reactions $^{93}$Nb(γ,$x$n; $x$ = 1 - 5)$^{(93-x)}$Nb and $^{100}$Mo(γ,n)$^{99}$Mo for monochromatic photons was carried out within the framework of the TALYS1.9 code with parameters by default {constant temperature model [42] and Fermi gas model [43]} in order to describe from a unified standpoint the energy dependence of the cross-sections of the observed reactions in the considered energy range of γ-quanta. The TALYS1.9 code allows calculating the cross-sections for photonuclear reactions in the range of photon energies up to 200

MeV. Subsequently, the values of the calculated cross-sections σ($E$) were weighted according to the bremsstrahlung flux in the energy range from the threshold of the corresponding reaction to the maximum energy of γ-quanta $E_{\gamma\max}$ = 33 ÷ 93 MeV. As a result of this procedure, the average cross-sections <σ($E_{\gamma\max}$)> were obtained according to the formula:

$$\langle\sigma(E_{\gamma\max})\rangle = \Phi^{-1}(E_{\gamma\max})\int_{E_{th}}^{E_{\gamma\max}}\sigma(E)W(E,E_{\gamma\max})dE\,, \tag{1}$$

where σ($E$) is the reaction cross-sections; $W(E, E_{\gamma\max})$ is the spectrum of bremsstrahlung photons with the upper limit $E_{\gamma\max}$; $\Phi(E_{\gamma\max}) = \int_{E_{th}}^{E_{\gamma\max}} W(E, E_{\gamma\max})dE$ is the integrated bremsstrahlung flux in the energy range from the reaction threshold $E_{th}$ to the maximum energy $E_{\gamma\max}$. The values <σ($E_{\gamma\max}$)> calculated in this way were compared with the experimental cross-sections, which were determined from the expression:

$$\langle\sigma(E_{\gamma\max})\rangle = \frac{\lambda\Delta A}{N_x I_\gamma \varepsilon\, \Phi(E_{\gamma\max})(1-e^{-\lambda t_{irr}})e^{-\lambda t_{cool}}(1-e^{-\lambda t_{meas}})}, \tag{2}$$

where Δ$A$ is the number of counts of γ-quanta in the full absorption peak (for the γ-line of the investigated reaction); $N_x$ is the number of target atoms; $I_\gamma$ is the absolute intensity of the analyzed γ-line; ε is the absolute detection efficiency for the analyzed γ-quanta energy; λ = ln2/$T_{1/2}$ is the decay constant.

In the present experiment, we measured the average cross-sections of photoneutron reactions with the formation of final nuclei in the isomeric (metastable) <σ($E_{\gamma\max}$)>$_m$ and ground states <σ($E_{\gamma\max}$)>$_g$. In addition, on the basis of these data, the average total cross-sections <σ($E_{\gamma\max}$)>$_{tot}$ = <σ($E_{\gamma\max}$)>$_m$ + <σ($E_{\gamma\max}$)>$_g$ were calculated, as well as the isomeric ratios of the average cross-sections of the reaction products (γ,4n) and (γ,5n):

$$d(E_{\gamma\max}) = \frac{\langle\sigma(E_{\gamma\max})\rangle_m}{\langle\sigma(E_{\gamma\max})\rangle_g} \tag{3}$$

Attention should be paid to the fact that the quantity <σ($E_{\gamma\max}$)> is a function that depends not only on the energy of γ-quanta, but also on the shape of the spectrum of bremsstrahlung γ-radiation, which will almost always be different due to the difference in experimental conditions in different laboratories. In this regard, in order to analyze the results of our experiment, the data obtained were compared with the values of <σ($E_{\gamma\max}$)> calculated by Eq. (1) for the bremsstrahlung spectrum from the converter used by us. At the same time, the results of other laboratories were compared with the cross-sections <σ($E_{\gamma\max}$)> calculated using a thin converter. In other words, when interpreting the experimental cross-sections <σ($E_{\gamma\max}$)>, the dynamics of their behavior with respect to the cross-sections curves calculated with TALYS1.9 using the shapes of the bremsstrahlung spectrum for the conditions of the corresponding experiment was considered.

**E. Calculation of photon fluxes and normalization by molybdenum activation**

The shape of the γ-ray bremsstrahlung spectrum was calculated using the GEANT4 code [44]. The calculations took into account the spatial and energy spread in the electron beam, as well as the distortion of the spectrum of bremsstrahlung photons introduced by the Al-absorber. The Al-absorber was also considered as an additional source of low-energy bremsstrahlung γ-radiation. We also calculated the bremsstrahlung spectrum of γ-quanta for a thin Ta-converter ($l$ ~ 50 μm). Such spectra are close in shape to the Schiff spectra [45] and were used to compare our results with other experiments performed on thin bremsstrahlung converters.

To normalize the bremsstrahlung flux, the reaction $^{100}$Mo(γ,n)$^{99}$Mo was used. The point is that calculations using the TALYS1.9 code reproduce with good accuracy the cross-sections of the (γ,n)-reaction on $^{100}$Mo in the range of γ-ray energies from the reaction threshold $E_{th}$ to 100 MeV. This, in turn, makes it possible to reliably calculate the average reaction cross-sections in the studied range of γ-quantum energies. Thus, comparing the experimental values of $\langle\sigma(E_{\gamma max})\rangle$ with the calculated ones, it is possible to estimate the possible deviation of the flux of bremsstrahlung γ-quanta calculated in the GEANT4 code from the real flux incident on the target.

In this regard, the calculated values of the cross-sections for the reaction $^{100}$Mo(γ,n)$^{99}$Mo, obtained using the TALYS1.9 code, were compared with the known data [46] in the GDR energy range ($E_\gamma$ = 8 ÷ 27 MeV). The analysis included the cross-sections represented by the Lorentz function:

$$\sigma(E) = \sigma_{max} \frac{(E\Gamma)^2}{(E^2 - E_{max}^2)^2 + E^2\Gamma^2}, \qquad (4)$$

where $\sigma_{max}$ is the cross-sections at maximum; $E_{max}$ and Γ are the position and width of the GDR. The values of these parameters were taken from Ref. [46], and were also fitted by us to the experimental data from Ref. [46] using the least square method. Figure 3(a) shows the experimental values of the cross-sections for the reaction $^{100}$Mo(γ,n)$^{99}$Mo in the energy range from 8 to 27 MeV [46], together with their representations by lines calculated using the TALYS1.9 code, the Lorentz function with parameters from Ref. [46], and the Lorentz function with parameters free.

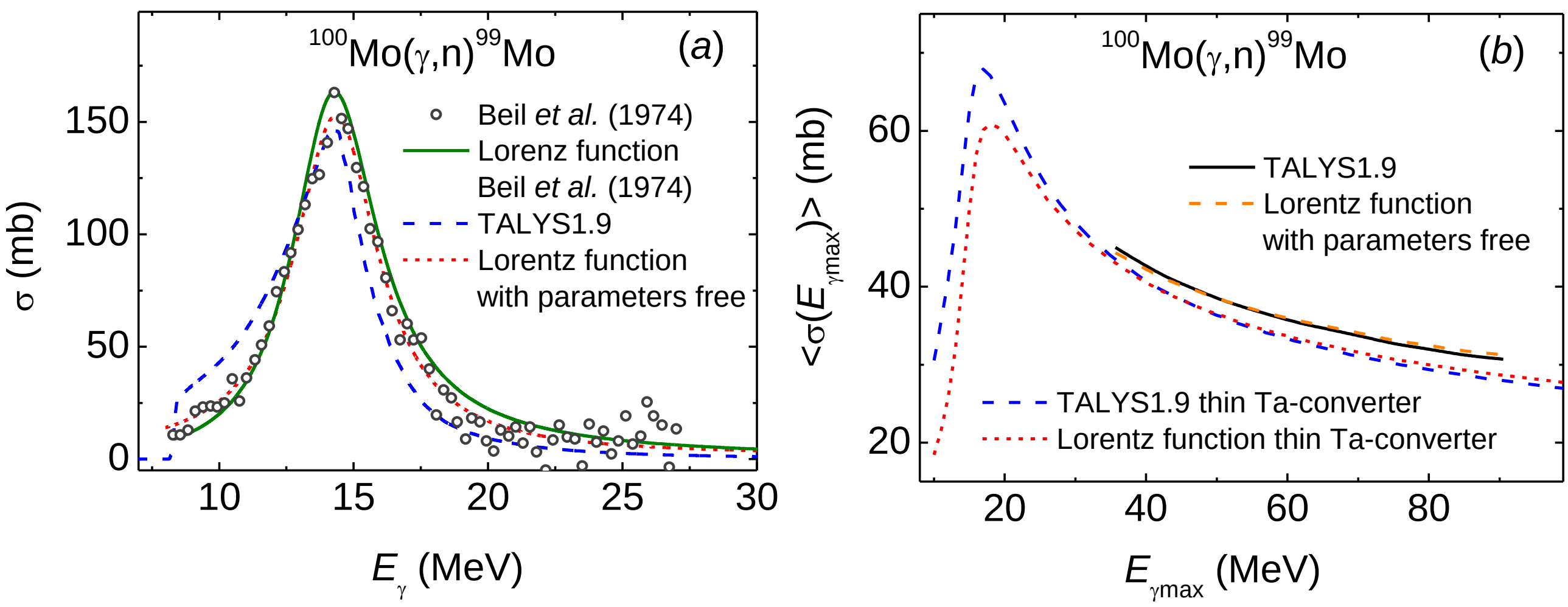

FIG. 3. (a) The experimental values of the cross-sections for the reaction $^{100}$Mo(γ,n)$^{99}$Mo in the energy range from 8 to 27 MeV [46], together with their representations by lines calculated using the TALYS1.9 code, the Lorentz function with parameters from Ref. [46], and the Lorentz function with free parameters. (b) The average cross-sections for the reaction $^{100}$Mo(γ,n)$^{99}$Mo calculated using the TALYS1.9 code and the Lorentz function with free parameters in the energy range $E_{\gamma max}$ = 33 ÷ 93 MeV both for the flux of bremsstrahlung γ-quanta of the present experiment and for the thin Ta-converter.

In the first two cases, the calculated cross-sections describe the experiment, but it seems that the calculation using the TALYS1.9 code gives slightly underestimated values in comparison with the experiment, and the Lorentz curve from Ref. [46] gives overestimated values. At the same time, the Lorentz function with free parameters describes the experiment [46] quite well, and its further extrapolation makes it possible to reliably calculate the average cross-sections in the energy range up to 100 MeV. Figure 3(b) shows that the average cross-sections calculated using the TALYS1.9 code and the Lorentz function with free parameters in the energy range $E_{\gamma max}$ = 33 ÷ 93 MeV differ insignificantly (~ 1 - 2 %) both for the flux of bremsstrahlung γ-quanta of the present experiment and for the thin Ta-converter. Thus, the calculation using the TALYS1.9 the average cross-sections of the $^{100}$Mo(γ,n)$^{99}$Mo reaction, which can be used to monitor the flux of bremsstrahlung γ-quanta.

In the $^{100}$Mo(γ,n)$^{99}$Mo reaction, the $^{99g}$Mo nucleus is formed only in the ground state $J^{\pi}$ = 1/2$^{+}$ with a half-life $T_{1/2}$ = 65.94±0.01 h. The ground state of $^{99g}$Mo undergoes $\beta^{-}$-decay with the formation of a $^{99}$Tc nucleus with excitation of several levels at $E_{\gamma}$ = 140.5 - 1198.9 keV, during the discharge of which two most feature γ-lines are observed with $E_{\gamma}$ = 140.51 and 739.50 keV with an absolute intensity $I_{\gamma}$ = 89.43 and 12.13%, respectively (see Table 1). Based on the values of $\Delta A$ for the indicated γ-lines, the experimental values of the average cross-sections $\langle\sigma(E_{\gamma max})\rangle_{exp}$ of the reaction under consideration were determined. Then, from a comparison of the experimental $\langle\sigma(E_{\gamma max})\rangle_{exp}$ and calculated $\langle\sigma(E_{\gamma max})\rangle_{th}$ values of the average cross-sections, the normalization coefficients $k = \langle\sigma(E_{\gamma max})\rangle_{th}/\langle\sigma(E_{\gamma max})\rangle_{exp}$ were obtained, which reflect the deviation of the flux of bremsstrahlung γ-quanta calculated in the GEANT4 code from the real flux that fell on the target. The found values of $k$, which varied in a rather narrow range of 1.08 ÷ 1.15, were subsequently used to correct the bremsstrahlung flux, which is necessary in determining the experimental and calculated values of the average cross-sections for photon-neutron reactions on the $^{93}$Nb nucleus.

The measurement error of the experimental values of the averaged cross-sections was determined as the quadratic sum of statistical and systematic errors. For different reactions $^{93}$Nb(γ,$x$n; $x$ = 1 - 5)$^{(93-x)}$Nb, both types of errors are different. The statistical error is mainly associated with the calculation of statistics in the peak of total absorption of the corresponding γ-line, which is estimated and varies within 1 - 10%. This error depends on the intensity of a particular γ-line and the background conditions for measuring the γ-spectrum. The intensity of the γ-line depends on the detection efficiency of the detector, the half-life $T_{1/2}$, and the absolute intensity $I_{\gamma}$. The γ-background is mainly determined by the contribution of the Compton scattering of high-energy γ-quanta.

Systematic errors are associated with the uncertainties of the irradiation time 0.5%, the electron current 0.5%, the detection efficiency of the detector ~ 3%, which is mainly associated with the error of the reference sources of .-radiation, the half-life of the reaction products and the absolute intensity of γ-rays 0-10%, which is noted in Table I, normalization of the experimental data for $^{93}$Nb to the $^{100}$Mo(γ,n)$^{99}$Mo monitor reaction yield 0.5-2%, calculation of the flux of bremsstrahlung photons in GEANT4 ~ 1.5%.

The total error in measuring the average cross-sections was 3 to 15%.

## III. RESULTS AND DISCUSSION

### A. $^{93}$Nb(γ,n)$^{92}$Nb reaction

As a result of the reaction $^{93}$Nb(γ,n)$^{92}$Nb, the final nucleus can be formed both in the ground state $J^{\pi}$ =7$^{+}$ and in the isomeric state $J^{\pi}$ =2$^{+}$ with half-lives $T_{1/2}$ = (3.47±0.24)×10$^{7}$ y and 10.15±0.02 d, respectively. The isomeric state of the $^{92}$Nb nucleus at $E^{*}$ = 135.5 keV undergoes $\beta^{+}$(EC capture)-decay with the excitation of five levels of the $^{92}$Zr nucleus, with the cascade γ-decay of which a characteristic γ-line $E_{\gamma}$ = 934.49 keV is observed with an absolute intensity of 99%. The $^{92}$Nb nucleus in the ground state also undergoes $\beta^+$(EC capture)-decay with the formation of an excited $^{91}$Zr nucleus at $E_\gamma$ = 1495.47 keV with $J^\pi$ =4$^+$, which decays with the emission of two γ-lines with $E_\gamma$ = 561.03 and 934.49 keV with intensities of 100% each. However, it is not possible to measure the yield of the indicated γ-lines because of the rather long half-life of the $^{92}$Nb nucleus in the ground state. In this regard, in the present experiments, only the $^{93}$Nb(γ,n)$^{92m}$Nb reaction was investigated by the γ-quanta yield with $E_\gamma$ = 934.49 keV. The values of the average cross-sections $\langle\sigma(E_{\gamma max})\rangle_m$ of the reaction under study obtained on the basis of these data in the range of limiting energies of the bremsstrahlung spectra $E_{\gamma max}$ =33 ÷ 93 MeV, as well as the corresponding calculations performed using the codes TALYS1.9 and GEANT4 are shown in Fig. 4.

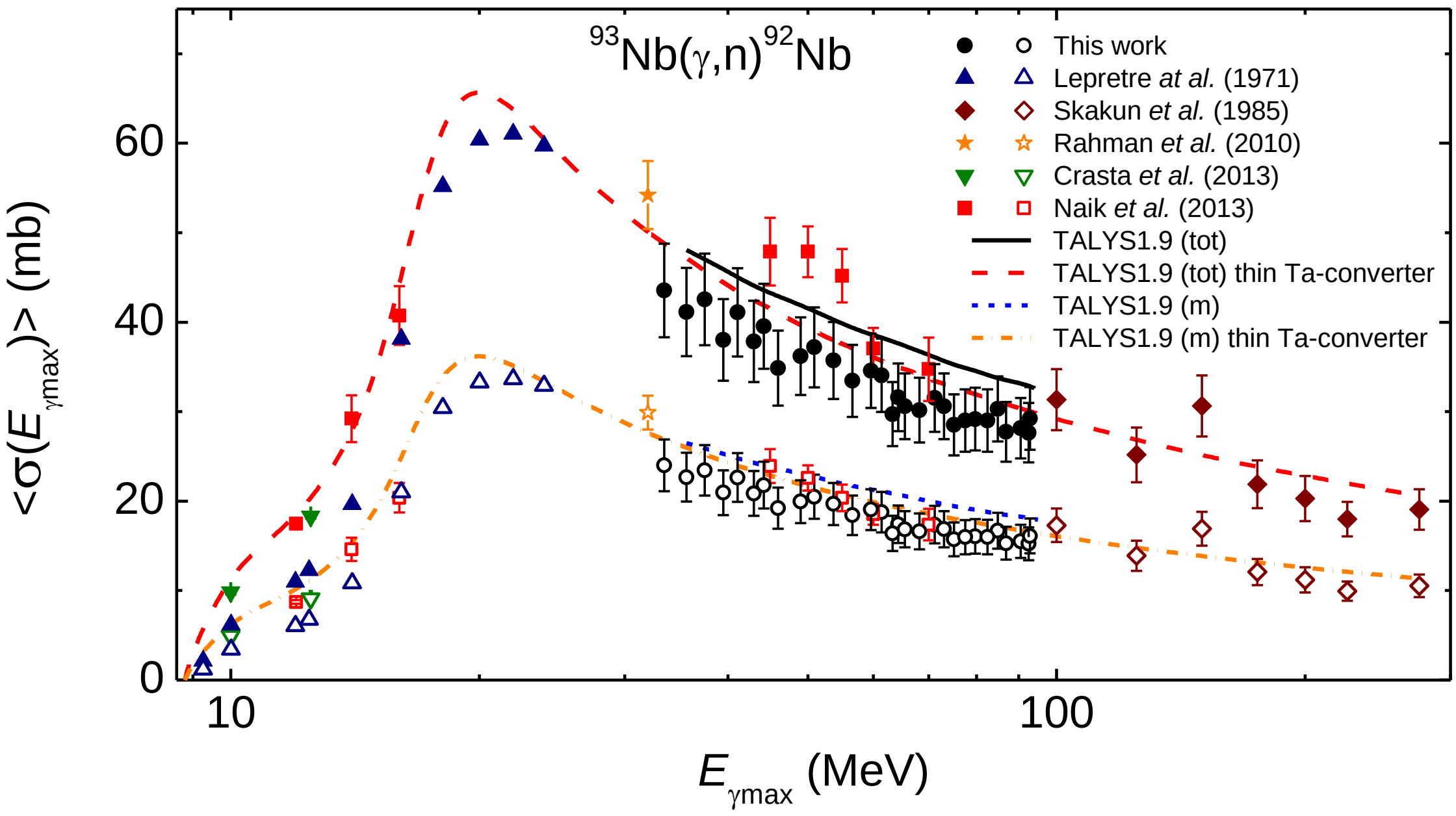

FIG. 4. Measured average cross-sections for the $^{93}$Nb(γ,n)$^{92}$Nb reaction compared with the previously published data together with theoretical calculations based on TALYS1.9 code. The experimental data are from Refs. [20-23, 27]. Filled symbols are for the total average cross-sections and open ones are for the $^{93}$Nb(γ,n)$^{92m}$Nb average cross-sections.

Figure 4 also shows the experimental values of the total cross-sections $\langle\sigma(E_{\gamma max})\rangle_{tot}$ of the reaction $^{93}$Nb(γ,n)$^{92}$Nb, which were obtained on the basis of the conclusions made in Refs. [23, 27]. For example, in the experiments by Skakun et al. [23], the value of the isomeric ratio of the average cross-sections per equivalent γ-quantum $\langle\sigma(E_{\gamma max})_Q\rangle$. for the formation of the $^{92}$Nb nucleus $d(E_{\gamma max}) = \langle\sigma(E_{\gamma max})_Q\rangle_m/\langle\sigma(E_{\gamma max})_Q\rangle_g \approx 1$ was estimated for the first time. As a confirmation, the authors compared the experimental values of the total cross-sections obtained by doubling the cross-sections $\langle\sigma(E_{\gamma max})_Q\rangle_m$ with the calculations performed within the framework of the cascade-evaporation model and obtained good agreement. In the studies of Rahman et al. [27], devoted to this issue, the contribution of the average cross-sections for the formation of a $^{92}$Nb nucleus in a metastable-state to the total average cross-sections of the (γ,n)-reaction at $E_{\gamma max}$ =32 MeV, which turned out to be equal to 55.2%, was experimentally determined, thereby confirming the conclusions of Ref. [23].

Indeed, this result is not unexpected and can be explained as follows. When dipole γ-quanta are absorbed by the $^{93}$Nb nucleus with a large spin $J^\pi$ = 9/2$^+$ of the ground state, states with $J$ = 11/2, 9/2, and 7/2 can be excited, the decay of which, taking into account the orbital angular momentum of the neutron, leads to comparable probabilities of formation of the final $^{92}$Nb nucleus in the isomeric and ground states. Our calculations using the TALYS1.9 code also showed that the ratio of the cross-sections for the formation of $^{92}$Nb in the isomeric state to the total cross-sections is 0.551. It was assumed that this estimate is valid in the entire energy range $E_{\gamma max}$ =33 ÷ 93 MeV and Fig. 4 shows the calculated data for the total cross-sections $\langle\sigma(E_{\gamma max})\rangle_{tot}$ of the reaction $^{93}$Nb(γ,n)$^{92}$Nb, and also the corresponding calculation of the cross-sections curve performed using the TALYS1.9 code. Figure 4 illustrates that the found values of the cross-sections $\langle\sigma(E_{\gamma max})\rangle_m$ and $\langle\sigma(E_{\gamma max})\rangle_{tot}$ turned out to be systematically lower than the theoretical calculations performed for the fluxes of bremsstrahlung γ-quanta of the present experiment. However, the results obtained are in satisfactory agreement with the data given by Naik et al. [22] for the reaction $^{93}$Nb(γ,n)$^{92m}$Nb, but deviate, within errors of up to 10%, for the total cross-sections at $E_{\gamma max} \leq 55$ MeV {the authors of Ref. [22] used the contribution coefficient of the cross-sections $\langle\sigma(E_{\gamma max})\rangle_m$ equal to 0.50}.

Satisfactory agreement is observed when comparing the $\langle\sigma(E_{\gamma max})\rangle_m$ cross-sections obtained on the basis of data from Skakun et al [23]. It should be noted that the experimental values of the cross-sections $\langle\sigma(E_{\gamma max})_Q\rangle_m$ given in Ref. [23] were obtained using a thin converter. For comparison with the calculations performed using the TALYS1.9 code, these data were recalculated to the values of the average cross-sections

$$\frac{\langle\sigma(E_{\gamma\max})\rangle}{\langle\sigma(E_{\gamma\max})_Q\rangle} = \frac{\int_0^{E_{\gamma\max}} W(E, E_{\gamma\max}) E dE}{\Phi(E_{\gamma\max}) \cdot E_{\gamma\max}}. \quad (5)$$

The corresponding total cross-sections were also estimated taking into account the contribution of the cross-sections $\langle\sigma(E_{\gamma max})\rangle_m$ with a coefficient of 0.551. The calculated values of the cross-sections $\langle\sigma(E_{\gamma max})\rangle_m$ and $\langle\sigma(E_{\gamma max})\rangle_{tot}$ for energies $E_{\gamma max}$ = 200 ÷ 275 MeV were obtained by extrapolating the curve calculated using the TALYS1.9 code to 200 MeV. As seen from Fig. 4, the experimental cross-sections of the $^{93}$Nb(γ,n)$^{92}$Nb reaction in the energy range $E_{\gamma max}$ = 100 ÷ 275 MeV demonstrate the same behavior dynamics as our results with respect to the theoretical values of $\langle\sigma(E_{\gamma max})\rangle$. On the whole, they also turned out to be systematically below the calculated curve. The same dynamics of the energy dependence can be traced for the average cross-sections in the region of the GDR maximum, which were obtained on the basis of detailed data by Lepretre et al. [20] on the cross-sections σ(*E*) of the (γ,n)-reaction. At the same time, the data from Ref. [27] at $E_{\gamma max}$ = 32 MeV turned out to be higher than the calculated values of $\langle\sigma(E_{\gamma max})\rangle_m$ and $\langle\sigma(E_{\gamma max})\rangle_{tot}$, while at $E_{\gamma max}$ = 10 and 12.5 MeV the data from Ref. [21] coincided with the calculation.

### B. $^{93}$Nb(γ,2n)$^{91}$Nb reaction

In the reaction (γ,2n) on $^{93}$Nb, the $^{91}$Nb daughter nucleus is formed in the ground and isomeric states with half-lives $T_{1/2}$ = 680±130 y and 60.86±0.22 d, respectively. In the ground state, the $^{91}$Nb nucleus undergoes $\beta^+$(EC)-decay into the $^{91}$Zr ground state with an intensity of 100%. In this regard, only the γ-decay of the $^{91m}$Nb isomeric state at $E^*$ = 104.49 keV with $J^\pi = 1/2^-$ was considered. The isomeric state decays by the internal *M*4-transition to the ground state $J^\pi = 9/2^+$ of the $^{91}$Nb nucleus with an intensity $I_\gamma$ = 0.539% and simultaneously undergoes $\beta^+$(EC)-decay into the $^{91}$Zr nucleus with excitation of the first level at $E_\gamma$ = 1204.77 keV with $J^\pi = 1/2^+$, which is discharged by the *E*2-transition to the ground state $J^\pi = 5/2^+$ with $I_\gamma$ = 2.96%. Thus, the yield of the $^{91m}$Nb product in the reaction (γ,2n) was identified by the γ-line with $E_\gamma$ = 1204.77 keV. Figure 5 shows the obtained experimental data on the average cross-sections for the reaction $^{93}$Nb(γ,2n)$^{91m}$Nb in the energy range $E_{\gamma max}$ = 33 ÷ 93 MeV and the corresponding calculation of the cross-sections curve performed using the TALYS1.9 code in the real geometry of the experiment.

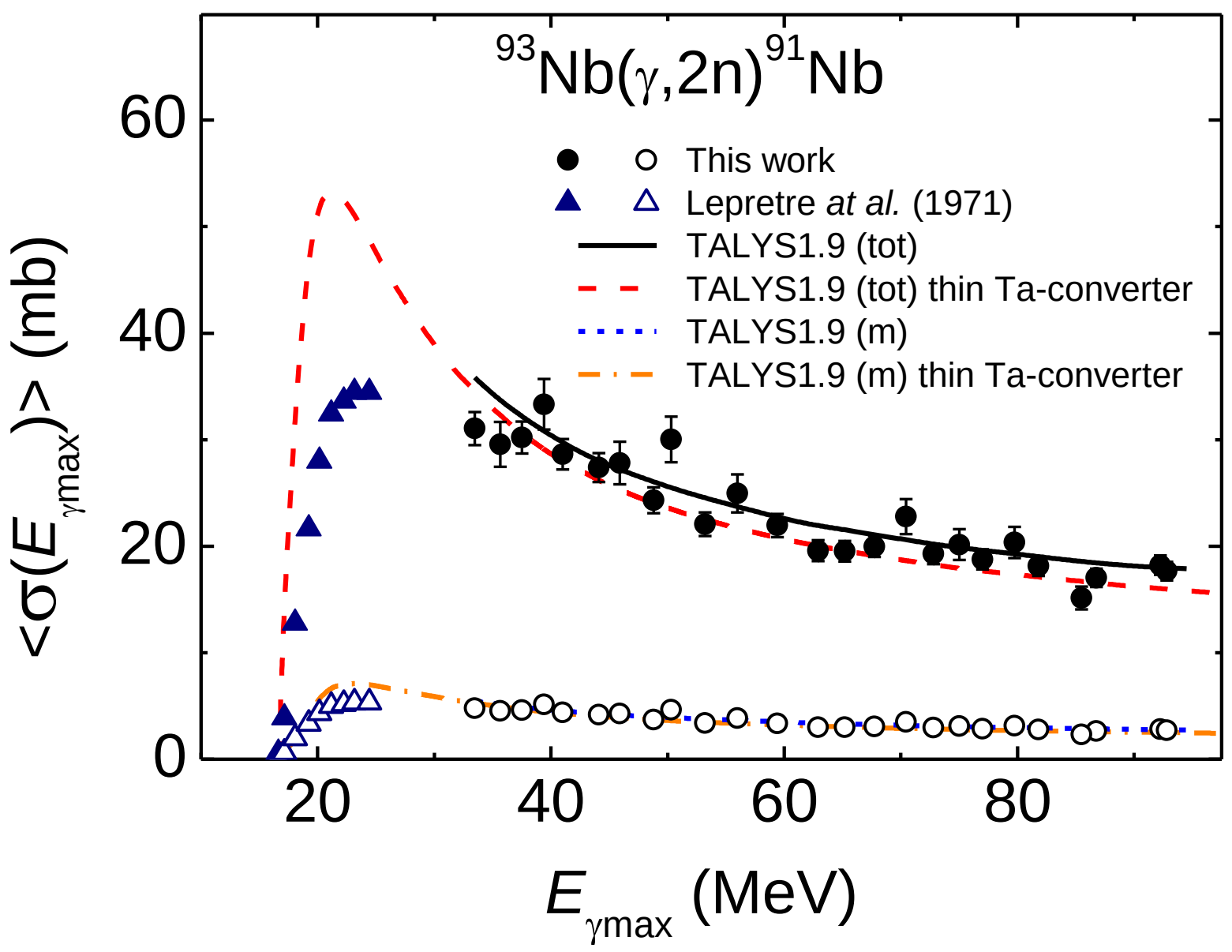

Fig. 5. Measured average cross-sections for the $^{93}$Nb(γ,2n)$^{91}$Nb reaction compared with the previously published data together with theoretical calculations based on TALYS1.9 code. The experimental data are from Ref. [20]. Filled symbols are for the total cross-sections and open ones are for the $^{93}$Nb(γ,2n)$^{91m}$Nb cross-sections.

As seen from Fig. 5, the experimental data on the cross-sections $\langle\sigma(E_{\gamma max})\rangle_m$ are in good agreement with the calculated values. From a comparison of the calculated average cross-sections for the formation of the $^{91}$Nb nucleus in the isomeric and ground states, a coefficient of 0.156 was obtained, which made it possible to estimate the possible experimental values of the total average cross-sections for the reaction $^{93}$Nb(γ,2n)$^{91}$Nb. Figure 5 shows that the estimate used led to reasonable values of the total average cross-sections for the reaction (γ,2n), despite the insignificant contribution of the cross-sections for the formation of the $^{91m}$Nb product nucleus to the total average cross-sections in the energy range $E_{\gamma max}$ = 33 ÷ 93 MeV. However, the obtained data on $\langle\sigma(E_{\gamma max})\rangle_{tot}$ turned out to be slightly lower than the calculated values. The same dynamics of the energy dependence is observed in the region of the GDR maximum, where detailed data on the cross-sections of the (γ,n)-reaction were obtained in the experiments of Lepretre et al. [20].

### C. $^{93}$Nb(γ,3n)$^{90}$Nb reaction

The $^{90}$Nb nucleus is formed in the ground and short-lived isomeric states with half-lives $T_{1/2}$ = 14.60±0.05 h and 18.81±0.06 s, respectively. The metastable-state at $E^*$ = 124.67 keV decays by an internal $E2$-transition to a level at $E^*$ = 122.370 keV, which is then discharged by an intense $M2$-transition to the ground state of the $^{90}$Nb nucleus with a branching ratio of 100%. Thus, the cross-sections for the formation of $^{90}$Nb in the ground state was considered as the total cross-sections for the reaction $^{93}$Nb(γ,3n)$^{90}$Nb. An unstable $^{90g}$Nb product nucleus was identified in the experimental spectrum (see Fig. 2) from the most intense γ-lines with $E_\gamma$ = 1129.23, 2186.24, and 2318.97 keV, observed during its decay. Figure 6 shows the experimental cross-sections $\langle\sigma(E_{\gamma max})\rangle_{tot}$ for the reaction $^{93}$Nb(γ,3n)$^{90}$Nb, weighted over the bremsstrahlung spectrum in the energy range from the reaction threshold to the maximum energy of γ-quanta $E_{\gamma max}$ = 33 ÷ 93 MeV and the corresponding calculation of the cross-sections curve executed using the TALYS1.9 code.

The experimental values of $\langle\sigma(E_{\gamma max})\rangle_{tot}$ in the region $E_{\gamma max}$ > 40 MeV systematically exceed the calculated ones predicted with the TALYS1.9 code. This disagreement between theory and experi-ment can be explained by the influence of shell effects, which can manifest themselves in connection with the filling of the $N = 50$ neutron shell in the $^{90}$Nb nucleus. For the reaction (γ,3n), the greatest competition is the channel (γ,p2n), which leads to the formation of a $^{90}$Zr nucleus with a filled neutron shell $N = 50$ and a proton subshell $Z = 40$, which has an underestimated level density.

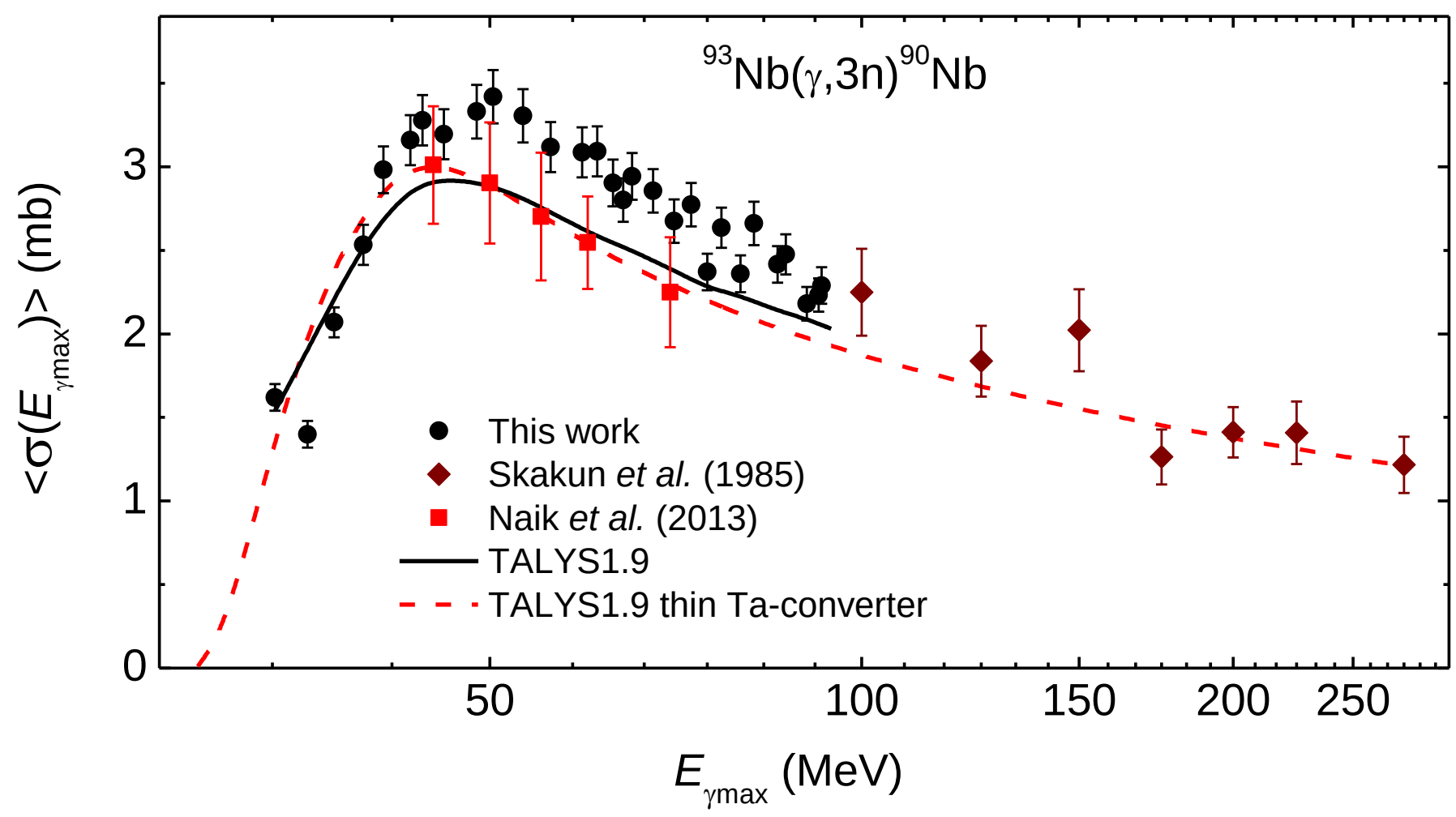

FIG. 6. Measured average total cross-sections for the $^{93}$Nb(γ,3n)$^{90}$Nb reaction compared with the previously published data together with theoretical calculations based on TALYS1.9 code. The experimental data are from Ref. [22, 23].

As seen from Fig. 6, our results, within the error limits, agree with the data given by Naik et al. [22] for the reaction $^{93}$Nb(γ,3n)$^{90}$Nb. However, the dynamics of the behavior of the reaction excitation function (γ,3n) from Ref. [22] with respect to the curve predicted using the TALYS1.9 code slightly differs from our results. At the same time, the data obtained by Skakun et al. [23] in the energy range at $E_{\gamma max} = 100 \div 275$ MeV, demonstrate the same behavior dynamics with our results with respect to the corresponding calculated cross-sections. Note that the experimental cross-sections for an equivalent quantum $\langle\sigma(E_{\gamma max})_Q\rangle_{tot}$ from Ref. [23], as in the case of the reaction $^{93}$Nb(γ,n)$^{92m}$Nb, were recalculated for the average cross-sections $\langle\sigma(E_{\gamma max})\rangle_{tot}$ using Eq. (5).

### D. $^{93}$Nb(γ,4n)$^{89}$Nb reaction

The $^{89}$Nb radionuclide is formed in the reaction (γ,4n) on $^{93}$Nb in the ground and isomeric states with half-lives of 1.9±0.2 h and 1.18±0.10 h, respectively. The ground state of the $^{89}$Nb nucleus undergoes $\beta^+$(EC)-decay with the excitation of several tens of levels of the $^{89}$Zr nucleus at $E_\gamma$ = 587.84 ÷ 3965.5 keV. When these levels are discharged, several γ-lines are observed with comparable absolute intensities. In this regard, for the identification of the $^{89g}$Nb radionuclide, γ-lines with $E_\gamma$ = 1511.7, 2572.3, 2960.1, and 3092.7 keV were used. The $^{92}$Nb nucleus in the isomeric state also undergoes $\beta^+$(EC)-decay with the formation of the $^{91}$Zr nucleus with excitation of 4 levels at $E_\gamma$ = 587.84 ÷ 1864.75 keV, during the decay of which three characteristic γ-lines are observed with $E_\gamma$ = 587.84 and 769.69 keV with intensities of 100 and 6.5%, respectively. Based on the data on the yields of the above mentioned γ-lines, the experimental values of the cross-sections $\langle\sigma(E_{\gamma max})\rangle_m$, $\langle\sigma(E_{\gamma max})\rangle_g$, and $\langle\sigma(E_{\gamma max})\rangle_{tot}$ for the reaction $^{93}$Nb(γ,4n)$^{89}$Nb, weighted over the bremsstrahlung spectrum in the energy range from the reaction threshold to the maximum energy of γ-quanta $E_{\gamma max}$ = 38.85 ÷ 93 MeV were determined. The results obtained, as well as the corresponding calculations using the TALYS1.9 code, are shown in Fig. 7.

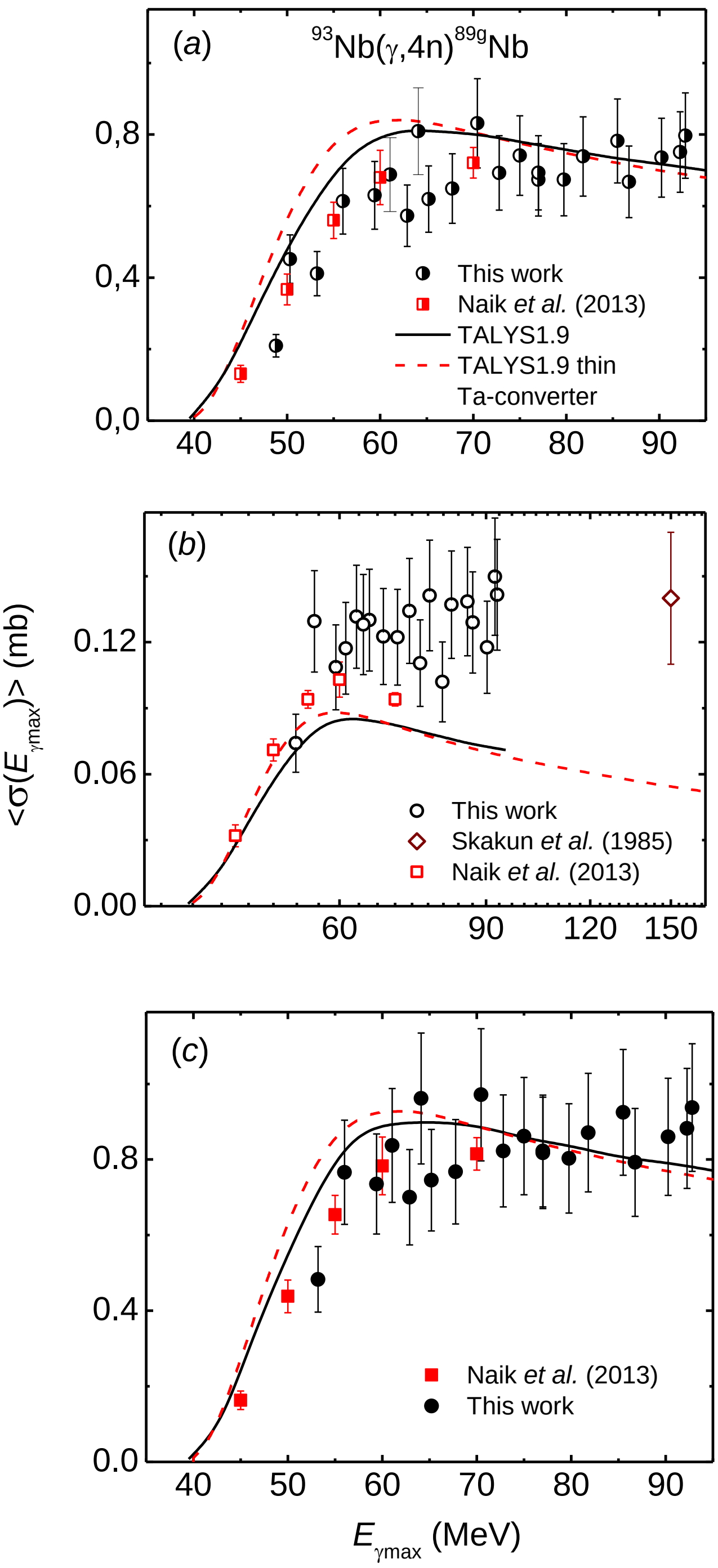

FIG.7. Measured average cross-sections for the $^{93}$Nb(γ,4n)$^{89}$Nb reaction compared with the previously published data together with theoretical calculations based on TALYS1.9 code. The experimental data are from Ref. [22, 23]. (a) $^{93}$Nb(γ,4n)$^{89g}$Nb. (b) $^{93}$Nb(γ,4n)$^{89m}$Nb. (c) The total cross-sections.

As seen from Fig. 7(a), the experimental data on the cross-sections $\langle\sigma(E_{\gamma max})\rangle_g$ of the reaction $^{93}$Nb(γ,4n)$^{89g}$Nb are in good agreement with the calculated values obtained using the TALYS1.9 code. The errors of the experimental data were mainly determined by the errors given for the values of the half-life and the absolute intensity of the γ-lines for the $^{89}$Nb product nucleus in the ground state (see Table 1). Our results for the reaction $^{93}$Nb(γ,4n)$^{89g}$Nb are in good agreement with the data given by Naik et al. [22].

The experimental data on the cross-sections $\langle\sigma(E_{\gamma max})\rangle_m$ of the reaction $^{93}$Nb(γ,4n)$^{89m}$Nb systematically exceed the calculated values predicted using the TALYS1.9 code [see Fig. 7(b)]. The same behavior dynamics is observed for the results obtained in Ref. [22], with which the present results are in good agreement. Our conclusions are also confirmed by the data [preliminarily recalculated according to Eq. (5)] obtained in Ref. [23] at $E_{\gamma max}$ = 150 MeV, which demonstrate the same behavior dynamics as our results with respect to the average cross-sections curve predicted using the TALYS1.9 code.

The experimental data on the total cross-sections $\langle\sigma(E_{\gamma max})\rangle_{tot}$ for the reaction $^{93}$Nb(γ,4n)$^{89}$Nb are in good agreement with the calculated values obtained using the TALYS1.9 code [see Fig. 7(c)]. This result is quite expected, since the contribution of the cross-sections for the formation of the $^{89}$Nb nucleus in the metastable-state to the total average cross-sections for the (γ,4n)-reaction is insignificant ~10%. Our results for the reaction $^{93}$Nb(γ,4n)$^{89}$Nb are in good agreement with the data given by Naik et al. [22].

### E. $^{93}$Nb(γ,5n)$^{88}$Nb reaction

The $^{88}$Nb nucleus is formed in the ground and isomeric states with half-lives $T_{1/2}$ = 7.8±0.1 min and 14.5±0.1 min, respectively. The isomeric state undergoes β$^+$(EC)-decay into a $^{91}$Zr nucleus with the excitation of several tens of levels at energies $E_\gamma$ = 1057.03 ÷ 4672.7 keV. The same situation is observed for the $^{88}$Nb nucleus in the ground state, which undergoes β$^+$(EC)-decay into the $^{88}$Zr nucleus with the excitation of several tens of levels at $E_\gamma$ = 1057.03 ÷ 5787.2 keV, when discharged, several γ-lines with the same energies are observed, as in the case of the decay of the isomeric state, with comparable intensities (see Table 1). In this regard, to identify the unstable nuclide $^{88}$Nb in the ground state, we used the “purest” γ-lines with $E_\gamma$ = 271.8 and 671.20 keV [see Fig. 2(a)], i.e., γ-lines that are not observed during the decay of the isomeric state. The same approach was used to identify the $^{93}$Nb(γ,5n)$^{88m}$Nb reaction channel, for which γ-lines with $E_\gamma$ = 450.5 and 760.8 keV were used. The obtained experimental values of the cross-sections $\langle\sigma(E_{\gamma max})\rangle_m$, $\langle\sigma(E_{\gamma max})\rangle_g$, and $\langle\sigma(E_{\gamma max})\rangle_{tot}$ for the reaction $^{93}$Nb(γ,5n)$^{88}$Nb, weighted over the bremsstrahlung spectrum in the energy range from the threshold reactions up to the maximum energy of γ-quanta $E_{\gamma max}$ = 51.41 ÷ 93 MeV, as well as their corresponding representations by lines calculated using the TALYS1.9 code are shown in Fig. 8.

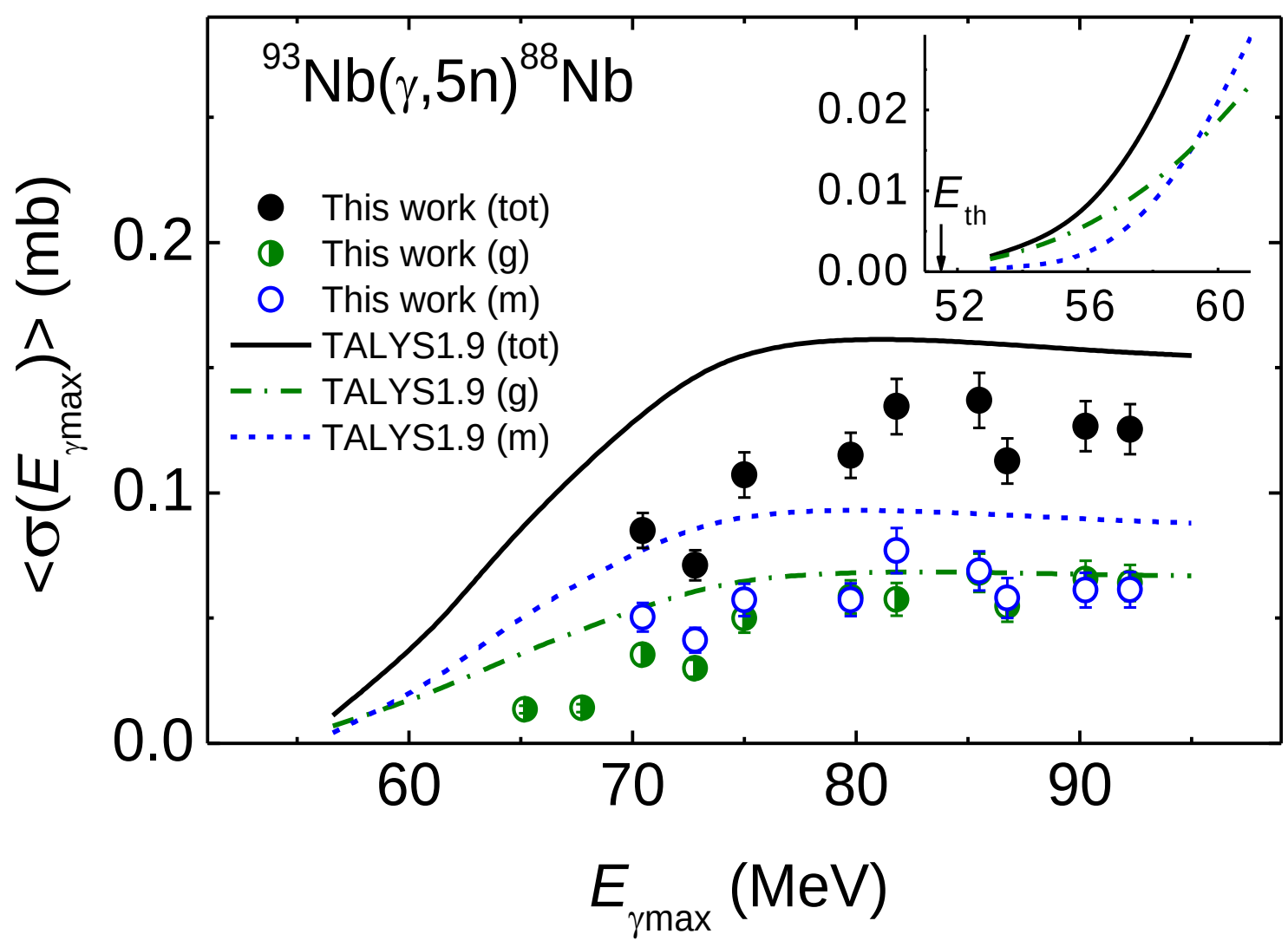

FIG. 8. Measured average cross-sections for the $^{93}$Nb(γ,5n)$^{88}$Nb reaction compared with theoretical calculations based on TALYS1.9 code. Inset - calculation of average cross-sections in the region of the threshold of the (γ,5n)-reaction.

The experimental data on the cross-sections <σ($E_{\gamma max}$)>$_g$ are in satisfactory agreement with the calculated values obtained using the TALYS1.9 code. A different situation is observed for the cross-sections <σ($E_{\gamma max}$)>$_m$, which practically coincided with the cross-sections for the formation of the $^{88g}$Nb nuclide in the ground state, and at the same time turned out to be two times lower than the calculated values of <σ($E_{\gamma max}$)>$_m$ predicted using the TALYS1.9 code. Naturally, this situation was reflected in the total cross-sections <σ($E_{\gamma max}$)>$_{tot}$ of the reaction $^{93}$Nb(γ,5n)$^{88}$Nb, which turned out to be lower than the calculated values predicted using the TALYS1.9 code. This disagreement between theory and experiment can be explained by the influence of shell effects, or at such high energies, competing channels associated with the emission of charged particles open. Additional and more detailed experiments would possibly confirm our statements about the dynamics of the behavior of the excitation functions of the reactions $^{93}$Nb(γ,5n)$^{89m}$Nb and $^{93}$Nb(γ,5n)$^{89g}$Nb with respect to their representations by lines calculated using the TALYS1.9 code.

At the end of this chapter, a summary Table II is given, which presents the cumulative measured values of the cross-sections <σ($E_{\gamma max}$)> and their uncertainties Δ<σ($E_{\gamma max}$)> for all investigated radionuclides formed in the corresponding photoneutron reactions on the $^{93}$Nb nucleus at $E_{\gamma max}$ = 33 ÷ 93 MeV.

TABLE II. Measured cross sections of Nb radio nuclei produced from the $^{93}$Nb(γ,*x*n; *x* = 1 - 5) reactions.

| $E_{\gamma max}$, [MeV] | ⟨σ($E_{\gamma max}$)⟩ ± Δ⟨σ($E_{\gamma max}$)⟩ (mb) | | | | | | |
|---|---|---|---|---|---|---|---|
| | $^{92m}$Nb | $^{91m}$Nb | $^{90}$Nb | $^{89m}$Nb | $^{89g}$Nb | $^{88m}$Nb | $^{88g}$Nb |
| 33.5 | 24.0±1.1 | 4.8±0.2 | 1.62±0.08 | | | | |
| 35.6 | 22.7±1.2 | 4.6±0.3 | 1.40±0.07 | | | | |
| 37.4 | 23.4±1.1 | 4.6±0.2 | 2.07±0.09 | | | | |
| 39.4 | 20.9±1.0 | 5.1±0.4 | 2.53±0.12 | | | | |
| 41.0 | 22.6±1.0 | 4.4±0.2 | 2.98±0.14 | | | | |
| 43.1 | 20.9±1.0 | - | 3.16±0.15 | | | | |
| 44.1 | 21.8±1.0 | 4.2±0.2 | 3.28±0.15 | | | | |
| 45.9 | 19.2±0.9 | 4.3±0.3 | 3.19±0.15 | | | | |
| 48.8 | 20.0±0.9 | 3.7±0.2 | 3.33±0.16 | | 0.21±0.03 | | |
| 50.3 | 20.5±0.9 | 4.6±0.3 | 3.42±0.16 | | 0.45±0.07 | | |
| 53.2 | 19.7±0.9 | 3.4±0.2 | 3.31±0.16 | 0.07±0.01 | 0.41±0.06 | | |
| 56.0 | 18.4±0.9 | 3.8±0.3 | 3.12±0.15 | 0.15±0.03 | 0.61±0.09 | | |
| 59.4 | 19.1±0.9 | 3.4±0.2 | 3.09±0.15 | 0.11±0.02 | 0.63±0.10 | | |
| 61.1 | 18.8±0.9 | - | 3.09±0.15 | 0.15±0.03 | 0.69±0.10 | | |
| 62.9 | 16.4±0.8 | 3.0±0.1 | 2.90±0.14 | 0.13±0.03 | 0.57±0.09 | | |
| 64.1 | 17.4±0.8 | - | 2.80±0.13 | 0.15±0.03 | 0.81±0.12 | | |
| 65.2 | 16.9±0.8 | 3.0±0.1 | 2.94±0.14 | 0.13±0.02 | 0.62±0.09 | - | 0.014±0.002 |
| 67.8 | 16.6±0.8 | 3.1±0.2 | 2.86±0.13 | 0.12±0.02 | 0.65±0.10 | - | 0.014±0.002 |

| $E_{\gamma max}$, [MeV] | ⟨σ($E_{\gamma max}$)⟩ ± Δ⟨σ($E_{\gamma max}$)⟩ (mb) | | | | | | |
|---|---|---|---|---|---|---|---|
| | $^{92m}$Nb | $^{91m}$Nb | $^{90}$Nb | $^{89m}$Nb | $^{89g}$Nb | $^{88m}$Nb | $^{88g}$Nb |
| 70.5 | 17.4±0.8 | 3.5±0.2 | 2.68±0.13 | 0.14±0.02 | 0.83±0.13 | 0.050±0.006 | 0.035±0.004 |
| 72.8 | 16.9±0.8 | 3.0±0.1 | 2.77±0.13 | 0.13±0.02 | 0.70±0.10 | 0,041±0,005 | 0.030±0.003 |
| 75.0 | 15.7±0.7 | 3.1±0.2 | 2.37±0.11 | 0.12±0.02 | 0.74±0.11 | 0.057±0.006 | 0.050±0.006 |
| 77.0 | 16.0±0.7 | 2.9±0.1 | 2.64±0.12 | 0.13±0.02 | 0.70±0.10 | - | - |
| 79.8 | 16.1±0.7 | 3.1±0.2 | 2.37±0.11 | 0.13±0.02 | 0.67±0.10 | 0.057±0.006 | 0.058±0.007 |
| 81.8 | 16.0±0.7 | 2.8±0.1 | 2.66±0.13 | 0.13±0,02 | 0.74±0.11 | 0.077±0.009 | 0.058±0.007 |
| 85.5 | 16.7±0.8 | 2.3±0.2 | 2.42±0.11 | 0.14±0.02 | 0.78±0.12 | 0.069±0.008 | 0.068±0.008 |
| 86.8 | 15.3±0.7 | 2.6±0.1 | 2.48±0.12 | 0,12±0.02 | 0.67±0.10 | 0.058±0.007 | 0.055±0.006 |
| 90.3 | 15.5±0.7 | - | 2.18±0.10 | 0.12±0.02 | 0.73±0.11 | 0.061±0.007 | 0.066±0.007 |
| 92.3 | 16.1±0.7 | 2.8±0.1 | 2.23±0.10 | 0.13±0.02 | 0.75±0.11 | 0.061±0.007 | 0.064±0.007 |
| 92.8 | 16.1±0.7 | 2.7±0.1 | 2.29±0.11 | 0.14±0.02 | 0.80±0.12 | - | - |

**F. Isomeric ratios**

Since in the experiment the values of the average cross-sections for the reaction products $^{93}$Nb(γ,4n)$^{89m,g}$Nb and $^{93}$Nb(γ,5n)$^{88m,g}$Nb were obtained, then Eq. (3) can be used to find their isomeric relationship. Figure 9 shows the experimental values of the isomeric ratios $d$ for (γ,4n)- and (γ,5n)-reactions. The character of their behavior as a function of the energy $d = f(E_{\gamma max})$ is not the same: for $^{89}$Nb it is a function smoothly decreasing from the threshold, reaching saturation in the energy region above 60 MeV, while for $^{88}$Nb this is, apparently, a function that sharply increases from the threshold and saturates in the energy region above 65 MeV. This behavior of the curves is due to the spins of metastable-states ($J^{\pi}$ =1/2$^-$ for $^{89m}$Nb and $J^{\pi}$ = 4$^-$ for $^{88m}$Nb).

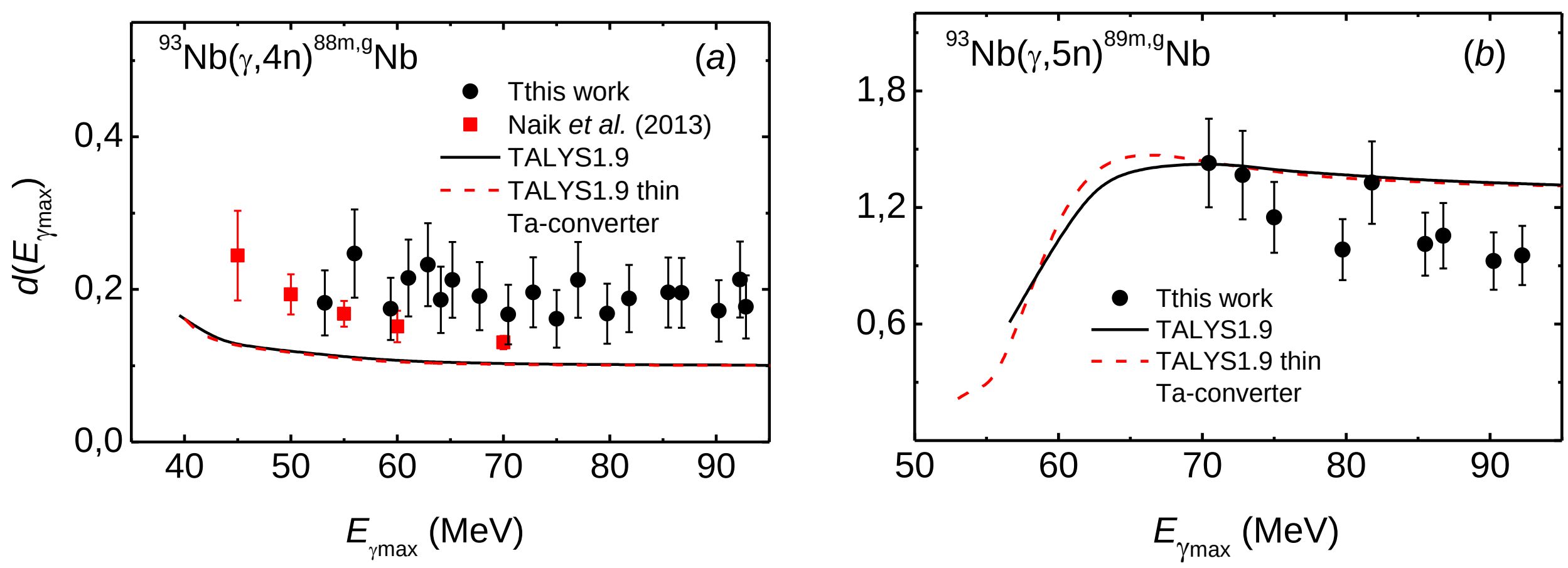

FIG. 9. Comparison of the measured isomeric ratios of the average cross-sections of the products of the $^{93}$Nb(γ,4n)$^{89m,g}$Nb (a) and $^{93}$Nb(γ,5n)$^{88m,g}$Nb (b) reactions with previously published data together with theoretical calculations based on the TALYS1.9 code. The experimental data are from Ref. [22].

As seen from Fig. 9(a), the expected values of isomeric ratios $d$ for the (γ,4n)-reaction, calculated in the framework of the TALYS1.9 code, are significantly lower, almost by a factor of 2, than the data of this experiment and the results from Ref. [22]. The errors in the data obtained were mainly determined by the errors given for the values of the half-life and absolute intensity of γ-lines for the $^{89}$Nb nucleus in the isomeric and ground states (see Table 1). More detailed calculations of isomeric ratios for this reaction were carried out by us earlier [32] using the TALYS1.8 code. In our simulations, we used several level density models incorporated in TALYS1.8, but we failed to satisfactorily describe the behavior of the energy dependence of isomeric ratios obtained in the experiments of Naik et al. [22].

For the (γ,5n)-reaction [see Fig. 9(b)], the calculated values of isomeric ratios are in satisfactory agreement with the experimental results. The general picture of the energy dependence of isomeric ratios for the (γ,5n)-reaction is understandable, despite the lack of data in the range of γ-quanta energies from the reaction threshold $E_{th}$ = 51.41 to 70 MeV. According to calculations performed using the TALYS1.9 code (see the inset in Fig. 8), the cross-sections for the formation of the $^{88g}$Nb nucleus significantly exceeds the cross-sections for the formation of the $^{88m}$Nb nucleus in the region of the reaction threshold $^{93}$Nb(γ,5n)$^{88}$Nb. However, with an increase in the energy of γ-quanta above 65 MeV, the situation changes dramatically - the cross-sections $\langle\sigma(E_{\gamma max})\rangle_m$ sharply increases relative to the value $\langle\sigma(E_{\gamma max})\rangle_g$. As a consequence, the isomeric ratios should have minimum values in the vicinity of the threshold of the (γ,5n)-reaction and, after 65 MeV, reach saturation. However, as noted in Subsection E, the data obtained for the cross-sections $\langle\sigma(E_{\gamma max})\rangle_g$ differ significantly from the calculated values, and, therefore, there is no reason to expect reasonable agreement between experiment and theory for isomeric ratios for (γ,5n)-reactions in the vicinity of the threshold. Perhaps more detailed experiments would confirm (or refute) our conclusions regarding the behavior of the energy dependence of isomeric ratios for the (γ,5n)-reaction in the energy range $E_{\gamma max} \approx 50 \div 70$ MeV.

Table III presents the cumulative experimental values of isomeric ratios for the reaction products $^{93}$Nb(γ,4n)$^{89m,g}$Nb at $E_{\gamma max} = 53 \div 93$ MeV and $^{93}$Nb(γ,5n)$^{88m,g}$Nb at $E_{\gamma max} = 70 \div 93$ MeV.

Table III. Measured isomeric ratios $d$ of the average cross sections of the products of the $^{93}$Nb(γ,4n)$^{89}$Nb and $^{93}$Nb(γ,5n)$^{88}$Nb reactions

| $E_{\gamma max}$, MeV | $d \pm \Delta d$ | |
|---|---|---|
| | $^{89m,g}$Nb | $^{88m,g}$Nb |
| 53.2 | 0.18±0.04 | |
| 56.0 | 0.25±0.06 | |
| 59.4 | 0.17±0.04 | |
| 61.1 | 0.21±0.05 | |
| 62.9 | 0.23±0.05 | |
| 64.1 | 0.19±0.04 | |
| 65.2 | 0.21±0.05 | |
| 67.8 | 0.19±0.04 | |
| 70.5 | 0.17±0.04 | 1.43±0.23 |
| 72.8 | 0.20±0.05 | 1.37±0.23 |

| $E_{\gamma max}$, MeV | $d \pm \Delta d$ | |
|---|---|---|
| | $^{89m,g}$Nb | $^{88m,g}$Nb |
| 75.0 | 0.16±0.04 | 1.15±0.18 |
| 77.0 | 0.21±0.05 | |
| 79.8 | 0.17±0.04 | 0.98±0.16 |
| 81.8 | 0.19±0.04 | 1.33±0.21 |
| 85.5 | 0.20±0.04 | 1.01±0.16 |
| 86.8 | 0.20±0.04 | 1.05±0.17 |
| 90.3 | 0.17±0.04 | |
| 92.3 | 0.21±0.05 | 0.95±0.15 |
| 92.8 | 0.17±0.04 | 0.92±0.15 |

## IV. CONCLUSIONS

In the present work, we measured the average cross-sections $\langle\sigma(E_{\gamma max})\rangle$ for the reactions $^{93}$Nb(γ,*x*n; *x* = 1 - 5 )$^{(93-x)m,g}$Nb in the energy range of the boundary energies of bremsstrahlung photons $E_{\gamma max}$ = 33 ÷ 93 MeV. The average cross-sections of the observed reactions were measured at 29 energies with a step $\Delta E_{\gamma max} \approx 2$ MeV. In addition, the isomeric ratios of the average cross-sections of the reaction products $^{93}$Nb(γ,4n)$^{89m,g}$Nb and $^{93}$Nb(γ,5n)$^{88m,g}$Nb were determined in the energy ranges $E_{\gamma max}$ = 53 ÷ 93 and 70 ÷ 93 MeV, respectively. Our data for the reactions (γ,n), (γ,3n), and (γ,4n) on $^{93}$Nb are in satisfactory agreement with the experimental results obtained by other authors at energies $E_{\gamma max}$ = 45, 50, 55, 60, and 70 MeV. The experimental cross-sections for the reactions (γ,2n) and (γ,5n) on $^{93}$Nb and the isomeric ratios of the reaction products $^{93}$Nb(γ,5n)$^{88m,g}$Nb were obtained for the first time.

Calculations of cross-sections, averaged cross-sections, and isomeric ratios have been performed using the TALYS1.9 code with default parameters. The measured average cross-sections and isomeric ratios were compared with theoretical calculations. For the reaction $^{93}$Nb(γ,5n)$^{88m,g}$Nb, which is characterized by the highest neutron emission, calculations in the TALYS1.9 code predict overestimated values of the averaged cross-sections than those obtained in the experiment. At the same time, for the reactions (γ,n) and (γ,3n), the calculation gives slightly underestimated values in comparison with the experiment. A possible reason for this deviation may be associated with the influence of shell effects, which are caused by the filling of the neutron shell in this mass region of the nuclei with $N$ = 50. On the other hand, for the reactions (γ,2n) and (γ,4n) on $^{93}$Nb, calculations in the TALYS1.9 code showed good agreement with the measured data. There is a tendency that, in the energy range under consideration, calculations using the TALYS1.9 code more successfully describe the average cross-sections of photoneutron reactions with the formation of final odd-even rather than odd-odd Nb nuclei. It can be concluded that the TALYS1.9 code with default parameters is an effective tool for evaluating the average cross-sections and isomeric ratios of photoneutron reactions at intermediate nuclei in the considered energy range of γ-quanta.

The new data obtained on the average cross-sections and isomeric ratios of the products of photon-neutron reactions on $^{93}$Nb will be important both for supplementing and updating nuclear databases, and can also be used for testing theoretical models used in modern computational codes.

## V. ACKNOWLEDGMENT

The authors would like to thank the crew of the linear electron accelerator LUE-40 SRE "Accelerator" NSC KIPT, Kharkov, Ukraine, for cooperation during the experiment.

---